\documentclass[12pt]{article}

\usepackage{amsmath, amsthm, amssymb}
\usepackage{color,soul}
\usepackage{graphicx}
\usepackage{epstopdf}
\usepackage{epsfig}
\usepackage{rotating}

\begin{document}

\title{Resolving Author Name Homonymy to Improve Resolution of Structures in Co-author Networks }

\author{Theresa Velden\\
{\small Department of Information Science, Cornell University} \\
{\small tav6@cornell.edu}\\
Asif-ul Haque\\
{\small Department of Computer Science, Cornell University}\\
{\small asif@cs.cornell.edu}\\
\ and
Carl Lagoze\\
{\small Department of Information Science, Cornell University}\\
{\small cjl2@cornell.edu}
}

\maketitle

\begin{abstract}
We investigate how author name homonymy distorts clustered large-scale co-author networks, and present a simple, effective, scalable and generalizable algorithm to ameliorate such distortions. We evaluate the performance of the algorithm to improve the resolution of mesoscopic network structures. To this end, we establish the ground truth for a sample of author names that is statistically representative of different types of nodes in the co-author network, distinguished by their role for the connectivity of the network. We finally observe that this distinction of node roles based on the mesoscopic structure of the network,  in combination with a quantification of author name commonality, suggests a new approach to assess network distortion by homonymy and to analyze the reduction of distortion in the network after disambiguation, without requiring ground truth sampling. 
\end{abstract}

\section{Introduction}


A nascent stream of research in scientometrics, policy research, and social studies of science and technology analyzes co-author or citation networks  to obtain a better understanding of scientific collaboration and the social organization of science. Author name ambiguity compromises this analysis and it is essential to remove this noise as the study of network structures becomes more sophisticated and moves beyond global measures of network topology to mesoscopic network features.  Whereas in the past, e.g. for the evaluation of scientists based on their publication output, manual disambiguation of author names was feasible,  large scale network studies require automated methods. 

We  present here a simple, effective, scalable and generalizable algorithmic approach for name disambiguation, and evaluate its performance in the particular use context of co-author network analysis. Based on our observations we suggest a new approach to assessing the quality of name disambiguation in co-author networks that does not require the expensive investment of establishing the ground truth for a representative sample, but builds exclusively on measures that can be derived from a structural analysis of the network itself. 

Name ambiguity can be classified into two kinds of problems: synonymy and homonymy. In this work we focus on name homonymy, which in the remainder of the paper we refer to as `name disambiguation'. In name homonymy, different individuals have the same name, either due to coincidence or abbreviations of names such as using initials for given names instead of using the full name. Homonymy is a problem especially for names coming from naming practices, such as those in  Korea or China, that may have uniquely identifiable full names but very common last names. 

Effective and generalizable author name disambiguation remains a generally unsolved problem for the following reasons. First,
different databases provide different kind of information about articles and authors (the feature-set used for disambiguation), making it hard to devise a 'one size fits all' algorithm. Second, the tolerance for errors  and for different types of errors will differ between use contexts. Third,  the methods for evaluating the effectiveness of a disambiguation algorithm are not well-established. No comprehensive, standardized set of benchmark data exists due to the variety in use contexts, the range of possibly relevant features of a dataset, and the costs of manually establishing ground truth.  Finally,  some algorithms do not scale for large data sets. All these concerns have resulted in a variety of algorithms for name disambiguation.

We  consider here the problem of name ambiguity in the context of earlier work
presented in~\cite{velden2010}. In this earlier work, we analyzed co-authorship
networks to better understand patterns of scientific collaboration in different scientific fields. We combined ethnographic methods with network analysis to identify co-author clusters in a co-author network as the smallest collective units of research in a field, and to extract linking patterns that represent different kinds of cooperative relationships between such collectives. A subnetwork of particular interest are the co-author clusters in a specialty field that show intensive inter-group collaboration. The work described here addresses the fact that this network, which was based on non-disambiguated author names,  showed peculiarly dense clustering for research groups with Asian affiliations, suggesting distortions due to name homonymy.

Our evaluation method is relatively novel compared to previous approaches because it takes network structural properties explicitly into account. We extract and quantify mesoscopic network features by classifying the nodes in a clustered co-author network into seven different classes of node roles based on their cluster internal and cluster external linking, following a classification scheme introduced by  \emph{Guimera et al.} in~\cite{guimera2007}. Given our suspicion of network distortions due to homonymy, we are interested to learn how those classes of nodes are affected by name homonymy and how their proportions change after disambiguation, reflecting changes in network structure. To establish the ground truth,  for each class of nodes we have sampled a representative set of author names and manually disambiguated them.  Based on this node role stratified sample, we can obtain estimates of the network distortions due to name homonymy, and we can evaluate the node role specific performance of our disambiguation algorithm.


Our algorithm for name disambiguation is fairly simple, yet effective, and can easily scale up for large networks. We consider two articles with
the same name to be by the same individual if either there is a co-author that is common
in both the articles, following an approach by \cite{kang2009}, or if there is a citation from one article to the other, which we
interpret as a self-citation. Co-author overlap is easy to compute and very effective,
while self-citation leverages an author's research continuity. One novel feature we
use in our algorithm is the commonality of last names, which we operationalize as author name redundancy by counting the number of variations of initials of a last name within our data set. Hence the distribution of name redundancy
can be easily obtained from the data set itself, and we present a principled
way of using this information for excluding less common names from unproductive disambiguation attempts. Also, through the use of name redundancy, those parts of the co-author network where naming traditions favor a small set of highly common last names become identifiable and quantifiable.


\section{Related Work}
There is a large body of work on name disambiguation which falls under the general area
of entity resolution (see~\cite{smallheiser2009} for a broad overview). These methods employ either supervised or unsupervised learning.


In supervised learning a smaller set of names is manually disambiguated so that a
classification model can be trained. In~\cite{han2004} techniques such as naive bayes and
support vector machines were employed effectively. The drawback of such methods is that
the training set needs to be large enough for the classifier to extrapolate unseen data accurately.
This re-introduces the problem of manual disambiguation of large sets of names.


Unsupervised learning uses clustering based on similarity metrics between names~\cite{han2005}.
Generative models such as latent dirichlet allocation and topic-based probabilistic latent
semantic indexing have also been used~\cite{bhattacharya2006, huang2006}.
The tricky part of using unsupervised learning is to judiciously choose the similarity metric and the
clustering algorithm. In~\cite{huang2006} the similarity metric was learned from a set of similarity
metrics via online active learning.


There are methods that tried to combine the benefits of supervised and unsupervised learning.
In~\cite{torvik2005, ferreira2010} training sets were generated automatically from the data.
Such training sets have noise in them and algorithms must not overfit by learning
the noise.


Whether learning is supervised or unsupervised, feature availability in the data and feature selection is of paramount importance.
Features regularly employed are co-author na\-mes, affiliation, article title, journal names and topic keywords~\cite{torvik2005, smallheiser2009, ferreira2010, han2004, han2005, huang2006, kang2009}. Unfortunately, affiliation on an author basis is not regularly available, nor are standardized keywords. Co-author names have been shown to be extremely effective~\cite{torvik2005, kang2009}, even by itself~\cite{kang2009}, and they provide a feature that is generally available in any data set of interest to author name disambiguation. Topics from article text were used in~\cite{song2007} while random walks on co-author networks were used in~\cite{malin2005}. An entirely different set of
features arises from reference or citation networks. For example, self-citation was used in~\cite{spencer2006}
and co-reference was used in~\cite{tang2010}. 


Because of overwhelming evidence in favor of co-author names we have chosen it as the main feature. We also use self-citation  to gain
more accuracy on top of co-author patterns. By using both co-author and citation based features we have broadened the grasp of our algorithm. One novel feature introduced in this work is
the quantification of the variety of first name initials associated with last names as an indicator of last name commonality. 

Our algorithm falls under the category of unsupervised learning where we have blocked the authors
by their names and clustered them using co-authorship and self-citation. It relies on clustering as simple as finding connected components on co-author overlap graphs, making it useable for large scale network analysis. The necessity for simplicity in large scale disambiguation was correctly noted in~\cite{smallheiser2009} and a recent attempt of disambiguation in the context of network analysis
was presented in~\cite{strotmann2010}. 

We do, however, have one parameter in our algorithm that was learned from a small set of manually disambiguated names. So our method is semi-supervised in some sense. But this parameter is based on a straightforward intuitive consideration, and the empirical determination mainly served to verify this intuition. We suggest that the learned value for this parameter can be safely applied to other data sets, so that our algorithm could be run in an unsupervised manner.

Because of the context of network analysis, our evaluation method is significantly different from previous works. Although name ambiguity is apparent in most standard bibliographic datasets, the importance or effect of disambiguating these authors is not apparent. In our evaluation we have taken into account the role of an author in a network and sampled authors from the seven roles (as presented in~\cite{guimera2007}) for manual disambiguation so that network structural effects of disambiguation can be assessed. 


\section{Data}

\begin{sidewaystable*}
\centering
\caption{Node Role Types \label{table:roles}}
\begin{tabular}{|c|c|c|c|c|c|c|} \hline
&Node&Characterization&Proportion in&Number in&Number in Ground-&Proportion of\\ 
&Role&&Population&Population&truth Sample&Population Sampled\\ \hline\hline
{\bf Non Hubs}&R1 &'ultra-peripheral nodes' & $30.3\%$&$5167$& $102$ &$1.97\%$\\  \cline{2-7}
&R2 &'peripheral nodes' &$48.4\%$&$8245$& $102$ &$1.24\%$  \\ \cline{2-7}
&R3 &'connector nodes' &$14.8\%$&$2527$& $102$ &$4.04\%$ \\  \cline{2-7}
&R4 &'satellite connector nodes' & $3.6\%$&$611$& $89$ &$14.57\%$\\ \hline\hline
{\bf Hubs}&R5 &'provincial hubs' &$1.1\%$&$195$& $72$ &$36.92\%$\\  \cline{2-7}
&R6 &'connector hubs' &$1.5\%$&$257$& $77$ &$29.96\%$ \\ \cline{2-7}
&R7 &'global hubs'  &$0.2\%$&$34$ & $28$ &$82.35\%$ \\ \hline
\end{tabular}
\end{sidewaystable*}

The publication data used in this study has been obtained from the Web of Science database by Thomson Reuters using a lexical query to capture the publications of a specialty field in physical chemistry over a period of 22 years (1987-2008). The co-author network constructed from this data set of $29,905$ publications, identifying individuals  based solely on first name initials and last name, was introduced in \cite{velden2010}. When building the co-author network we filter out and exclude from the network author names that have only one paper associated with them\footnote{This filtering is applied to reduce noise in the network structure of a scientific community. It excludes about 20\% of publications from the data set and 60\% of author names. The filtering is not perfect, as in a first step author names with only one publication get excluded, and then in a second step we remove all now orphaned publications, or publications with only one remaining author. This step may result in some authors now having only a single publication left in the data set. We interrupt the recursive process at this point, leaving about  $3\%$ of authors with a single publication included in the network.}, and end up with $18,419$ nodes, representing authors linked by co-authorship, with a giant component of $17,250$ nodes ($93.7\%$).

Clustering of the co-author network using the information theoretic clustering in~\cite{rosvall2007}, exposes the modular structure of co-author relationships, and results in a network of clusters of closely collaborating authors. Each author node in such a clustered network can be classified into one of seven node role types introduced in \cite{guimera2007}. A node is classified as a hub node or a non-hub node based on a first parameter, the number of its cluster internal links relative to the average inside-the-cluster degree of the nodes in the respective cluster.  This means a hub node in a cluster has more cluster internal links than the average node of that cluster. A second parameter quantifies how a node distributes its outside links among the clusters and subdivides hub nodes into three groups, and non-hub nodes into four groups, both of which are ordered by increasing outside linking. See table~\ref{table:roles} for characterizations of those type of nodes and their frequency in the giant component of our network. As reported in~\cite{velden2010}, based on this distinction between node roles, we can find a typical principal investigator (PI) led, hierarchically organized research group as a starlike structure, represented by a hub node in the center of a cluster with smaller nonhub nodes around, or a field-specific research institution or funded research network  as a more egalitarian organized cluster with several hub nodes involved.

In the following we focus on the giant component of the coauthor network, and population statistics are based on all nodes in the giant component that can be classified according to \emph{Guimera et al.'s} role type classification\footnote{For a few clusters zero standard deviation of the inside-the-cluster degree prevents calculation of the first parameter needed in the classification, resulting in the exclusion of $1.2\%$ of nodes in the giant component from the population.}. This population comprises $92.5\%$ of the nodes in the entire (undisambiguated) network. For this population at least $75\%$ of papers are published by coauthor teams of 5 or less authors (median 3, mean 3.8). The maximum number of coauthors found is $34$. 

As described below, the classification of author nodes is significantly distorted by author name homonymy, affecting in particular externally linking node role types (R3, R4, R6, R7). This is of concern; for the study of collaboration between groups, the resolution of nodes with role types characterized by high between-cluster linking is crucial, since they determine the connectivity of the inter-group collaboration network.

\subsection*{Name Redundancy}

\begin{figure}
\centering
\epsfig{file=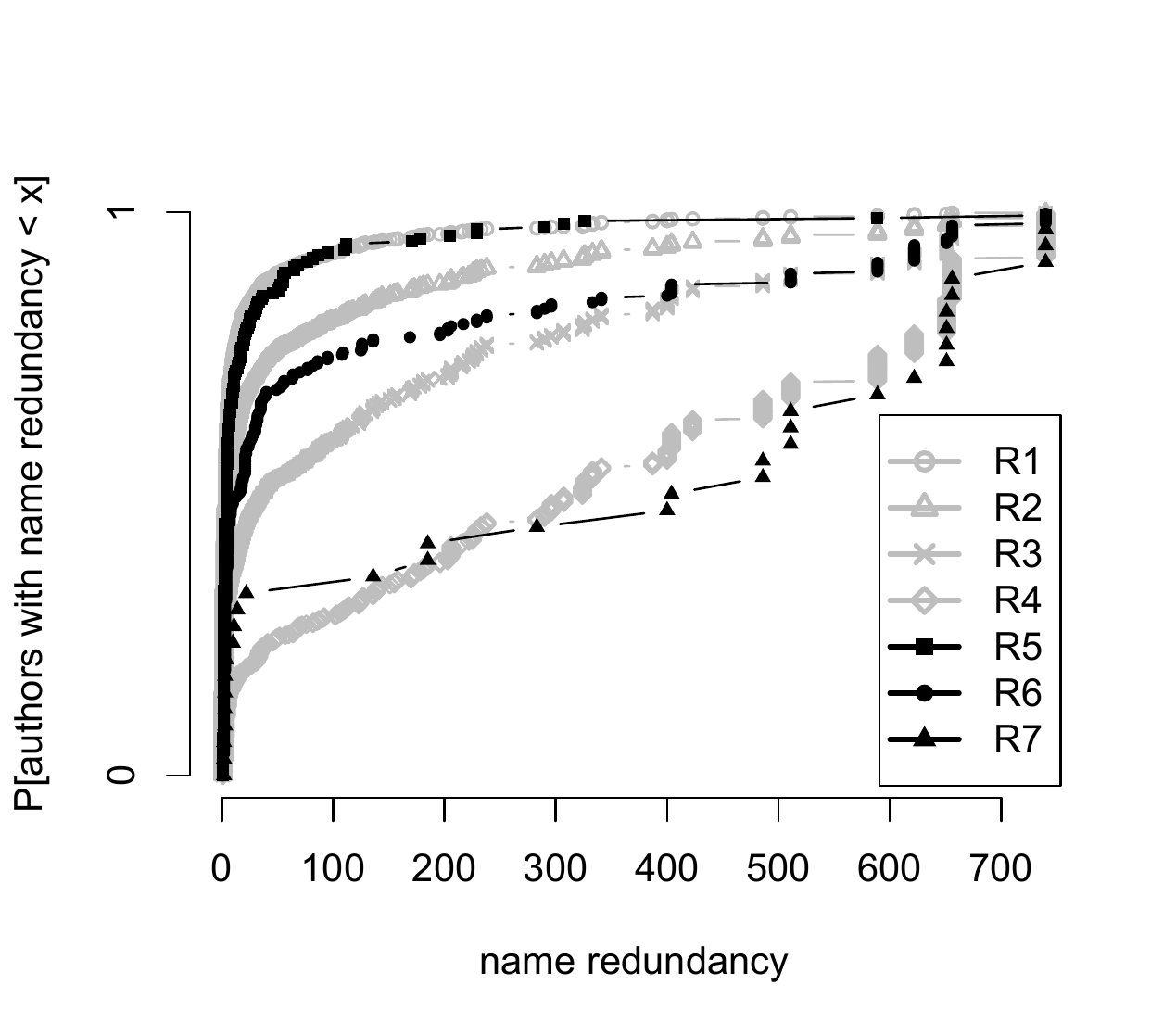, height= 10cm, width= 10cm}
\caption{Cumulative probability distributions of name redundancies 
before disambiguation.}
\label{fig:nameredundancies-nondis}
\end{figure}

To capture the ambiguity of an author name due to homonymy we introduce a measure of a name's commonality that we derive from the data set itself. We call it `raw name redundancy', and it is obtained by examining how numerous variations of initials with the same last name are. For example, for the Chinese last name `WANG"  we have found in our data set $740$ instances of names containing the last name ``WANG"  that can be distinguished by their initials, like ``WANG, CH'' can be distinguished from ``WANG, XL''. Another example for a high scoring last name is the Korean name ``LEE" with raw name redundancy of $511$. A large portion of the last names appearing in our data set, $91.7\%$, have raw name redundancy  of $3$ or less. It is worth noting though that of the $86 ,389$ co-authorship instances (an author being listed as a coauthor for a paper), $52,913$  ($61.2\%$) are attributed to authors with raw name redundancies greater than $3$, suggesting the larger number of actual authors represented by that smaller proportion of names. 

The comparison of the cumulative probabilities of raw name redundancy for the seven different node role types in our population data set in  fig~\ref{fig:nameredundancies-nondis} reveals strongly right-skewed, long-tailed distributions for role types R1, R2 and R5, that become heavy-tailed for R3 and R6, have an even distribution for R7, and are finally-left skewed for R4. This indicates an overrepresentation of very high redundancy names for the externally linking node role types R3, R4, R6 and R7. We suggest that one would expect randomness in the distribution of raw name redundancies among node role types, resulting in very similar curves. That the curves differ significantly suggests that a substantial portion of these nodes has been wrongly classified as strongly outward connecting, due to a lack of resolution of distinct author identities because of name homonymy. This would imply a serious misrepresentation of the true interconnectivity between co-author clusters in the undisambiguated network.

If we observe a last name $L$ to have $r_n(L)$ different initials associated with it in the dataset then we define its ``name redundancy'' $s_n(L)$ to be the cumulative normalized $r_n(L)$ value: \begin{displaymath} s_n(L) = \text{\bf Pr}[X \le r_n(L)] \end{displaymath} Here $X$ is the random variable on $r_n(.)$ distribution, and $r_n(L)$ the ``raw redundancy'' of $L$. Last names with small raw redundancy will have name redundancy close to 0 while last names with many different initials will score close to 1.

\begin{figure}
\centering
\epsfig{file=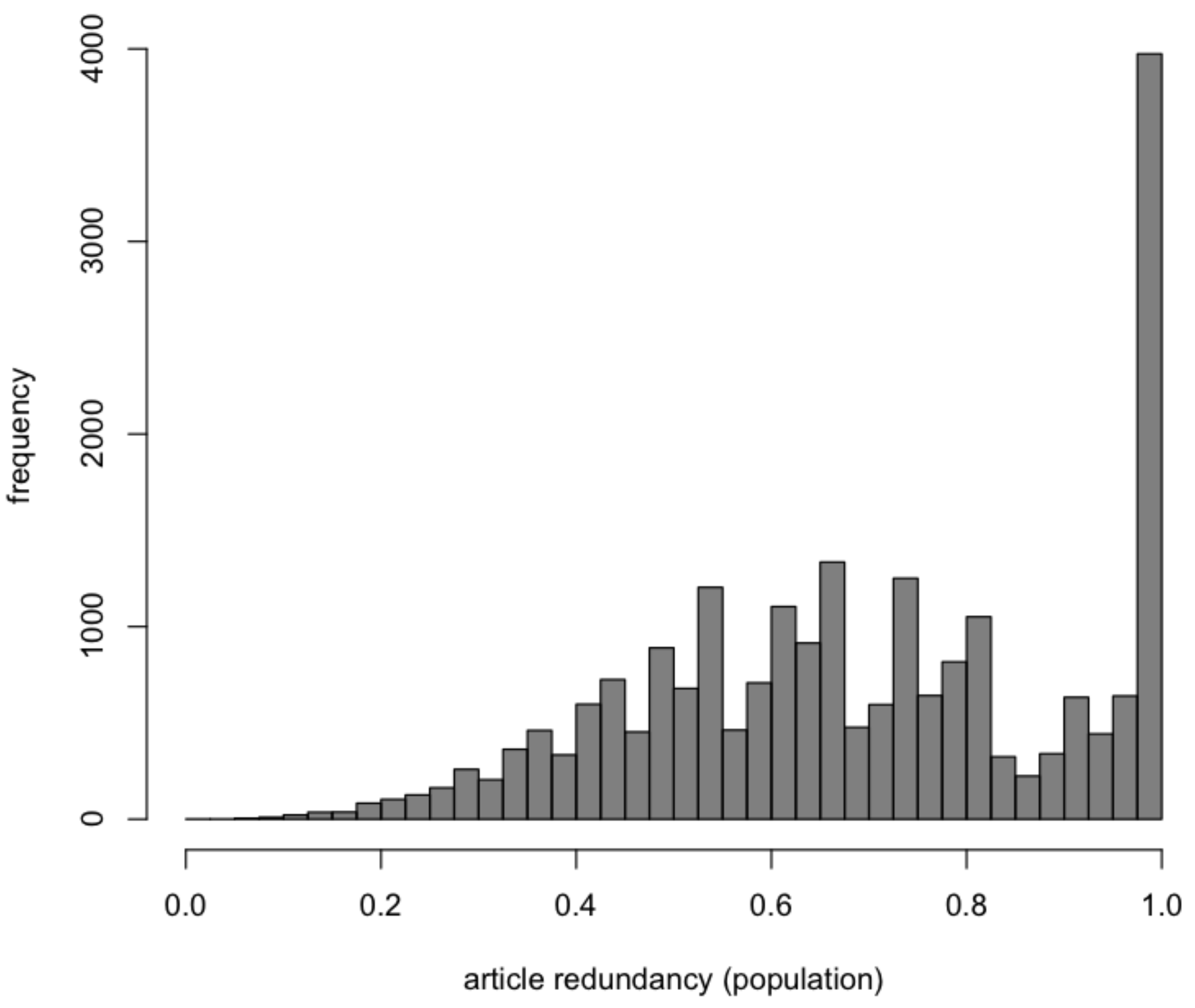, height= 6cm, width= 8cm}
\caption{Distribution of article redundancy for the population data set resulting from the combined name redundancies of the authors of an article.}
\label{fig:articleredundancies-population}
\end{figure}

Building on this definition we introduce as ``article redundancy" the combined name redundancies of the co-author team writing an article, defined as the product of name redundancies of the last names of the authors. The  distribution of article redundancies for the articles of the authors included in our population data set shows two distinct regions, one symmetric broad distribution, and one narrow peak, fig~\ref{fig:articleredundancies-population}. Those can be conceptualized as the overlap of two distributions. The broad distribution comprises articles with author teams that include one or several author last names with low name redundancy. Assuming an average number of co-authors per paper of roughly four authors, this distribution would result from the 4-fold convolution of distributions representing the independent, random choice of last names from the name redundancy distribution.  The narrow peak on the other hand can be interpreted as the result of the convolution of distributions representing the independent choice of last names exclusively from the heavy tail of the name redundancy distribution. Upon inspection of manually selected samples we conclude that these are mainly East Asian, specifically Chinese and Korean, last names. Hence we suggest that the shape of the distribution in this diagram highlights the division of our data set into two components that are culturally (naming traditions) and geographically (co-location of closely collaborating authors) distinct.

To check how this division in the article set is reflected in the set of authors, we can calculate for each author name the average article redundancy for all articles authored by an author of that name. We find that the distribution of average article redundancies in the population data set (represented by the combination of white and light grey areas of the bars in fig~\ref{fig:averagearticleredundancy-comparison}) also shows a distinct peak for high average article redundancies that result if the author name and all the names of the co-authors of that author have high name redundancy scores. Such region in the average article redundancy distribution is of much interest since they are most affected by homonymy, as shown further below. We can derive a rough estimate from the distribution in fig~\ref{fig:averagearticleredundancy-comparison} on the proportion of authors represented by the high average article redundancy peak, and hence the percentage of authors working in exclusive teams of coauthors with highly common last names, likely geographically co-located in East Asian countries. Taking the local minimum at about 0.85 as the division point, we find $21.5\%$ of (non-disambiguated) authors in the population data set belong to the peak of high ($> 0.85$) average article redundancy. 

\subsection*{Ground Truth}
\label{sec:groundtruth}
To estimate the error made by not correcting for homonymy in author names, and to quantify the improvement made by our disambiguation approach, we randomly sample a subset of $571$ author names from the population for manual disambiguation of author identities. The procedure for establishing the ground truth for this sample is described in the appendix.  
 
To account for systematic differences between the different node role types, we stratified the sample by node role type and sized the sample strata to be able to make statements on sample proportions  with at least a confidence interval of $10\%$, and a level of confidence of $95\%$. Sample sizes are reported in the two rightmost columns in table~\ref{table:roles}. We sampled an additional $33\%$ of author names for each groundtruth stratum. This provided us with a training set for verifying our intuition about a low-name-redundancy cut-off parameter that excludes extremely uncommon names from any disambiguation attempt in order to avoid introducing unnecessary error (see respective subsection in sec.~\ref{parameters}).

\begin{figure}
\centering
\epsfig{file=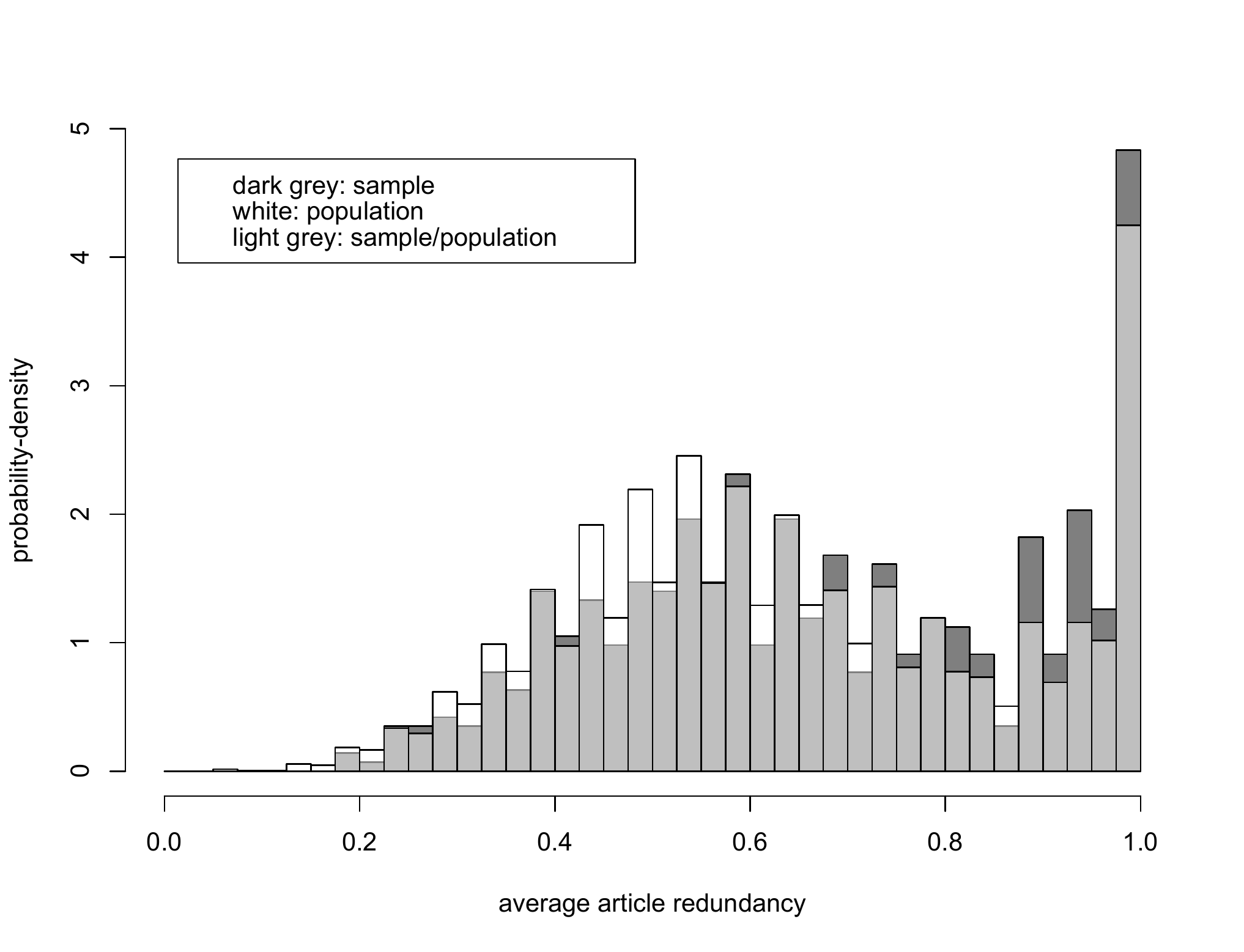, height= 6cm, width= 8cm}
\caption{Probability density distribution of average article redundancy for author names.}
 \label{fig:averagearticleredundancy-comparison}
\end{figure}

Note that the groundtruth sample when aggregated across the node role strata does not reflect the actual proportions of node role types in the population (shown in the rightmost column in table~\ref{table:roles}), simply because their relative proportions in the ground truth sample are not representative for their relative proportion in the population\footnote{This is a consequence of choosing the minimal necessary sample size for each stratum because establishing ground truth is an expensive time-consuming process. Those minimal strata sizes have been calculated following p.131 in~ \cite{Rea1997}, and those numbers fall off significantly for large population sizes. Hence, in order to have a groundtruth sample representative of the relative node role proportions in the population, one would have to match the required minimal sampling number for the least frequent node role type R7, by sampling e.g. for the most frequent node role type R2, $275$ author names instead of the $102$ author names included in our ground truth sample.}. Consequently,  when interpreting results for the aggregate groundtruth set one has to keep in mind that  one can make straightforward statistical estimates only within each stratum, i.e. for a specific node role type. 
The comparison of the distribution of the average article redundancy of the sample with that of the population, as depicted in fig~\ref{fig:averagearticleredundancy-comparison}, shows that the sample aggregate is biased toward higher average article redundancies, a reflection of the fact that node role types most affected by name redundancy are overrepresented in the ground truth sample.

\section{Algorithm}


The basic idea of our algorithm is simple: two papers authored by an author with the same name are
highly likely to be works of the same author if the two papers share common
co-authors. Following~\cite{kang2009} we use overlap of two coauthor sets by at least one last name as sufficient to merge two author identities. The result is the growth of connected components in co-author overlap graphs. 

Furthermore, if a paper cites another, and both papers are authored by an author with the same name, then very often this a is  self-citation reflecting the research continuity of that author. Although weaker than co-authorship, we have found signal
from self-citation to be very accurate.

Finally, we have found that authors with last names that are unique in our data set are best disambiguated by considering every
occurrence of such a name as referring to the same individual. The most uncommon last names will show up in our data set with raw name redundancy of $1$. Intuitively,  because often the same name is written with last name plus $2$ or $3$ different variations of initials, such as first initial, first and middle initials, or solely middle initial, we might want to include names with raw name redundancies of $2$ or even $3$ into that set of 'unique' names.

We do not use affiliation and city information when available in our dataset since it is difficult 
to associate those with authors in a principled manner. We also do not use any text or 
topic content such as title, journal or keywords because of our dataset being from a 
narrow subfield of chemistry. These features may be discriminative for a large heterogeneous
dataset like PubMed, but are less useful for a narrow research area where a lot of articles share the
same keywords and are published in a few journals. We have investigated the applicability of
tf-idf similarity of the abstracts and it indeed turned out to be less informative.

Thus we use in our method of disambiguation co-authors and self-citation on those names whose
redundancy is beyond a certain value that we call the low redundancy
cut-off, which we determine from the training data set to verify our intuition.

\subsection*{K Metric}

The ground truth specifies for a set of articles with the same author name subgroupings or clusters of articles,  each cluster for a different individual with that author name. In order to compare this 'true' clustering with either the trivial clustering for the undisambiguated data (all papers with the same author name form one group) or with the clustering resulting from an automated disambiguation attempt, we need a measure of the agreement between those  clusterings. The accuracy of a clustering with respect to the
true clustering, can be quantified in a number of different ways. The metric we found most 
relevant is the ``K metric'' used in \cite{ferreira2010}. Given the true
clusters for a name there are two quantities of interest for an empirical
clustering: the average cluster purity (ACP) and the average author purity (AAP).
Cluster purity is high when an empirical cluster contains articles mostly by
the same individual. But cluster purity does not quantify how fragmented
a cluster is. In the extreme case a true cluster may be split into many
singleton clusters, each with high cluster purity. Author purity quantifies
the correctness of the splits. For a true cluster if all the articles are in
the same empirical cluster the author purity is perfect. The K metric combines
the cluster and author purities. It is defined as the geometric mean of the
average cluster purity and the average author purity.

For a name let there be $N$ articles ($N$ nodes in the article graph constructed by our algorithm)
which in reality represent $t$ individuals. Suppose the $j^\text{th}$ individual,
or cluster, contains $n_j$ articles. So $\sum_{j=1}^t n_j = N$. Suppose the
grouping of the same articles produced by our algorithm has $e$ clusters
where the $i^\text{th}$ cluster has $n_i$ articles. Thus $\sum_{i=1}^e n_i = N$. The \emph{average cluster purity}({\bf ACP}) and the \emph{average author purity}
({\bf AAP}) are defined as follows.
\begin{displaymath} \text{\bf ACP} = \frac{1}{N}\sum_{i=1}^e\sum_{j=1}^t \frac{n^2_{ij}}{n_i}
\end{displaymath}
\begin{displaymath} \text{\bf AAP} = \frac{1}{N}\sum_{j=1}^t\sum_{i=1}^e \frac{n^2_{ij}}{n_j}
\end{displaymath}
Here $n_{ij}$ is the number of articles that are in true cluster $j$ as well as in
empirical cluster $i$. So $\sum_{i=1}^e\sum_{j=1}^t n_{ij} = N$.
\begin{displaymath} \text{\bf K} = \sqrt{\text{\bf ACP} \times \text{\bf AAP}} \end{displaymath}

The K values for our data are widely distributed.  For this reason we have used quantiles in parameter learning and algorithm evaluation, rather than averages to aggregate the K distributions. Further, we have weighted the distribution of K values with the size of the article set for each names since this size is indicative of the importance of disambiguating that name.

\subsection*{Parameter Learning}
\label{parameters}
\begin{figure}
\centering
\epsfig{file=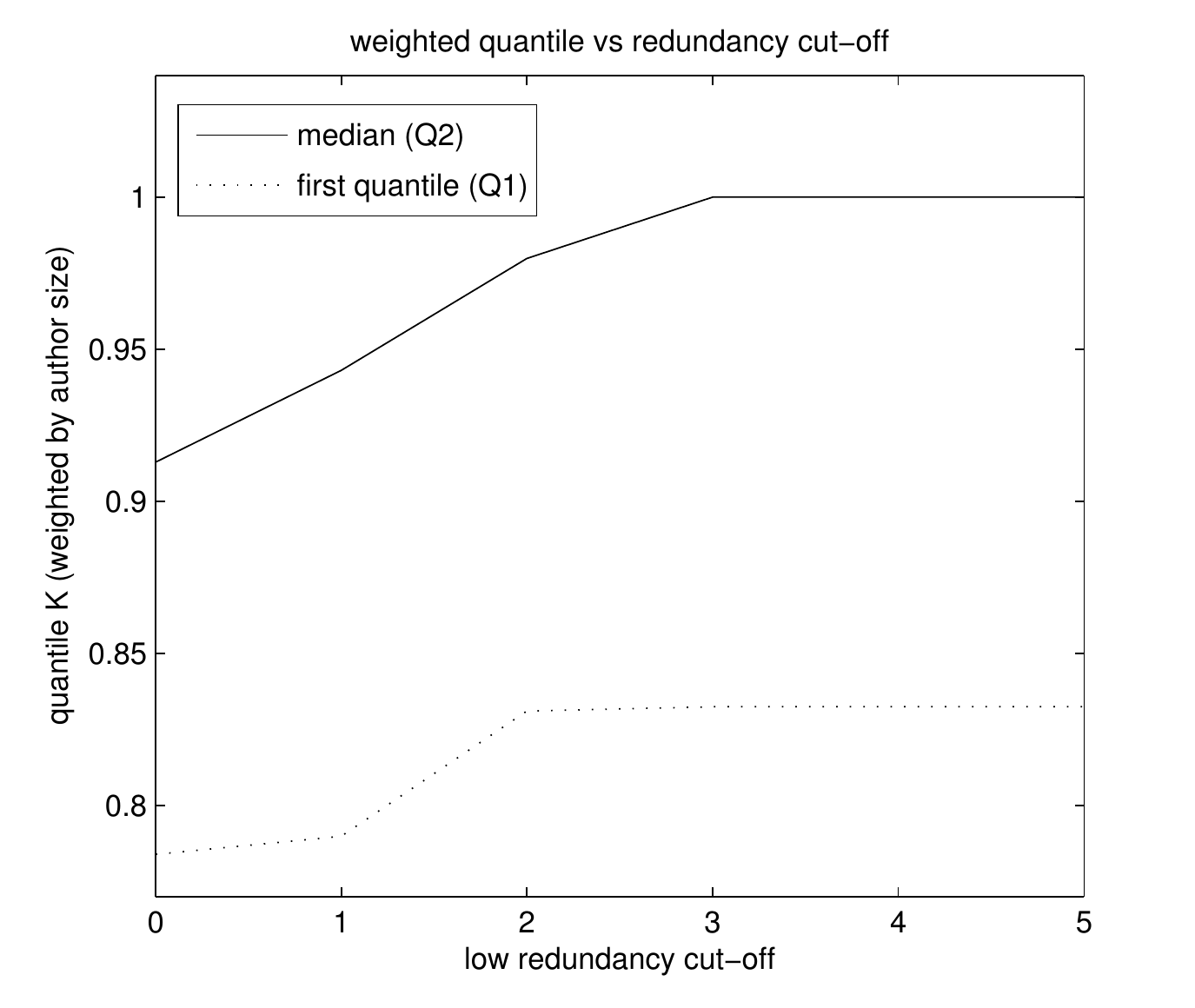, height= 6cm, width= 8cm}
\caption{Quantiles of weighted K values for the training authors for each low redundancy cut-off.}
\label{fig:lowredquantile}
\end{figure}

Our disambiguation algorithm has one parameter, the low redundancy cut-off. Last names with redundancy scores below this threshold are assumed to refer to the same individual. This parameter was learned from the training set of author names without using self-citation information. The result of a series of runs with different low name redundancy cut-off on the training data is shown in
fig~\ref{fig:lowredquantile}. For each cut-off value, last names with raw redundancy less than or equal to it were trivially disambiguated by considering each of them to be one single identity. For last names above the cut-off, co-author overlap was used for disambiguation. A cut-off value of zero meant all names were disambiguated via co-author overlap. The weighted median K curve in fig~\ref{fig:lowredquantile} shows $3$ to be the best low redundancy cut-off value. For the lower end of the K distribution, $3$ is also the optimal cut-off as shown by the weighted first quantile in fig~\ref{fig:lowredquantile}. This confirms our intuition that a name with such low raw redundancy is better disambiguated by merging all appearances of the name.

\section{Results}

In this section we present what we learned from the ground truth data, namely the extent of node distortions due to author name homonymy. We derive an estimate for the error rate in the entire population, which comprises almost all of the nodes in the giant component of the undisambiguated co-author network. We also use the ground truth data to measure the improvements made by using our version of the disambiguation algorithm. In addition we discuss changes in network structure when moving from the undisambiguated network to the disambiguated version, demonstrating how name disambiguation is critical for the resolution of mesoscopic network features. Finally, we report an observation that suggests a method to assess and compare the level of disambiguation in a coauthor network, without having to invest into the expensive creation of a ground truth data set. 

\begin{figure}
\centering
\epsfig{file=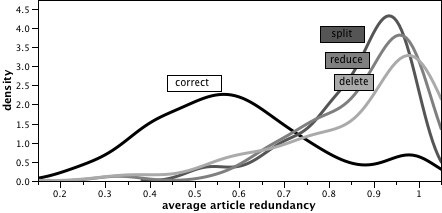, height= 4cm, width= 8cm}
\caption{Kernel density estimates for error types for undisambiguated author names in ground truth data set. \label{fig:resultypes}}
\end{figure}
\begin{table}
\centering
\caption{Role Specific Distortions by Homonymy \label{table:errors}}
\begin{tabular}{|l|c|c|c|c|c|c|c|} \hline
&R1&R2&R3&R4&R5&R6&R7\\ 
&[$\%$]&[$\%$]&[$\%$]&[$\%$]&[$\%$]&[$\%$]&[$\%$]\\ \hline
correct&$98.0$&$80.4$&$51.5$&$22.5$&$88.9$&$72.7$&$32.1$\\ \hline
reduce&$0$&$7.8$&$11.9$&$16.9$&$6.9$&$10.4$&$28.6$\\ \hline
split&$1.0$&$3.9$&$10.9$&$11.2$&$4.2$&$13.0$&$17.9.$\\ \hline
delete&$1.0$&$7.8$&$25.7$&$49.4$&$0$&$3.9$&$21.4$\\ \hline
\hline\end{tabular}
\end{table}
\subsection*{Distortions in Undisambiguated Network} 
Based on the true identity of authors established for the author names in the groundtruth data set, we can derive estimates for the errors made by not disambiguating author names for homonymy. We distinguish three error types to reflect the different effects correcting them would have on the actual nodes in the network.  'Split' means that the ground truth suggests that a node is split into at least two authors with a minimum of two papers each. 'Reduce' means a node is to be reduced in size since additional authors were found each of which has no more than one paper, and hence does not survive initial filtering of data when building the network. Finally 'delete' means a node is split into separate identities none of which has more than one paper, deleting the node entirely from the network, again due to filtering out of one-paper authors when building the network. 

Table~\ref{table:errors} shows, for the nodes in the ground truth data set, the different kinds of errors that were made by representing all instances of an author name by the same node, as if they all referred to the same individual. Based on these results we obtain the following estimates\footnote{Approximate error margins given for a $95\%$ confidence interval} of the proportion of correct nodes in the giant component of the non-disambiguated network: of the $R1$ non-hub nodes, almost all, $98\% (\pm 0)$ correctly represent a single author,  followed by the $R5$ hub nodes with $88.9\% (\pm 1.1)$ correctly representing a single author.  For R2 and R6 nodes the non-disambiguated network represents a large majority of nodes correctly,  with $80.4\%$ $(\pm 1.7)$, and $72.7.\%$$ (\pm 2.9)$, respectively. Those rates go dramatically down for R3, R7 and R4 nodes, with $51.5.\%$$ (\pm 4.6)$,  $32.1.\%$$ (\pm 12.5)$, and $22.5\%$$ (\pm 8.0)$ of nodes correctly representing a single author. 

These results confirm our suspicion that the issue of name homonymy causes misrepresentation of individual authors especially for those nodes that determine the inter-cluster connectivity of the clustered network. So, whereas the most numerous node role types in the network, R1 and R2, have small error rates, and the overall estimated error rate across all  node role types is about $20\%$, the error estimate for those nodes of role types that most crucially determine the mesoscopic structure of the collaboration network, those that link between clusters that represent research groups, rise to $68\%$, and $78\%$ for R7, and R4 nodes, respectively.  

Fig~\ref{fig:resultypes} indicates how the different error types are distributed over the range of average article redundancies of the author names. The density estimates of the various errors due to false merges peak for author names that have very high average article redundancies, i.e. authors publishing exclusively with colleagues that also have very common last names, leading to distortions in the respective parts of the co-author network.

\subsection*{Evaluation of Disambiguation Algorithm}

Table~\ref{table:quantiles} compares for author names in the groundtruth sample the weighted K quantiles before and after disambiguation. Results are reported for the node role specific strata of the sample. The median of weighted $K$ shows significant improvements after disambiguation for node roles R4 and R7, further improvements at the lower $25\%$ quantile level for R3, R4, and R7, and a slight decrease for R6 nodes. There are also significant improvements of the minimum values of the weighted $K$ distributions for all node role types, except R1 and R5. Excluding those author names that do not show any change in $K$, the median gains (or losses) in $K$ by node role type for those author names that did change are as follows: R1: $-0.2$,  R2: $0.25$,  R3: $0.28$,  R4: $0.42$, R5: $0.01$,  R6: $0.01$,  R7: $0.42$. 

\begin{table}
\centering
\caption{Quantiles of Weighted $K$\label{table:quantiles}}
\begin{tabular}{|l|c|c||c|c||c|c|} \hline
&\multicolumn{2}{|c||}{\bf median}&\multicolumn{2}{|c||}{\bf 25\%}&\multicolumn{2}{|c|}{\bf minimum}\\ \cline{2-7}
&nondis&dis&nondis&dis&nondis&dis\\ \hline  \hline
{\bf R1}&$1.0$&$1.0$&$1.0$&$1.0$&$0.71$&$0.61$   \\ \hline
{\bf R2}&$1.0$&$1.0$&$1.0$&$1.0$&$0.44$&$0.68$\\ \hline
{\bf R3}&$0.85$&$1.0$&$0.65$&$0.89$&$0.39$&$0.56$\\ \hline
{\bf R4}&$0.5$&$1.0$&$0.4$&$0.89$&$0.28$&$0.58$\\ \hline
{\bf R5}&$1.0$&$1.0$&$1.0$&$1.0$&$0.62$&$0.57$\\ \hline
{\bf R6}&$1.0$&$1.0$&$1.0$&$0.98$&$0.41$&$0.59$\\ \hline
{\bf R7}&$0.54$&$0.93$&$0.28$&$0.89$&$0.2$&$0.69$\\ \hline 
\end{tabular}
\end{table}

\begin{figure*}
\centering
\epsfig{file=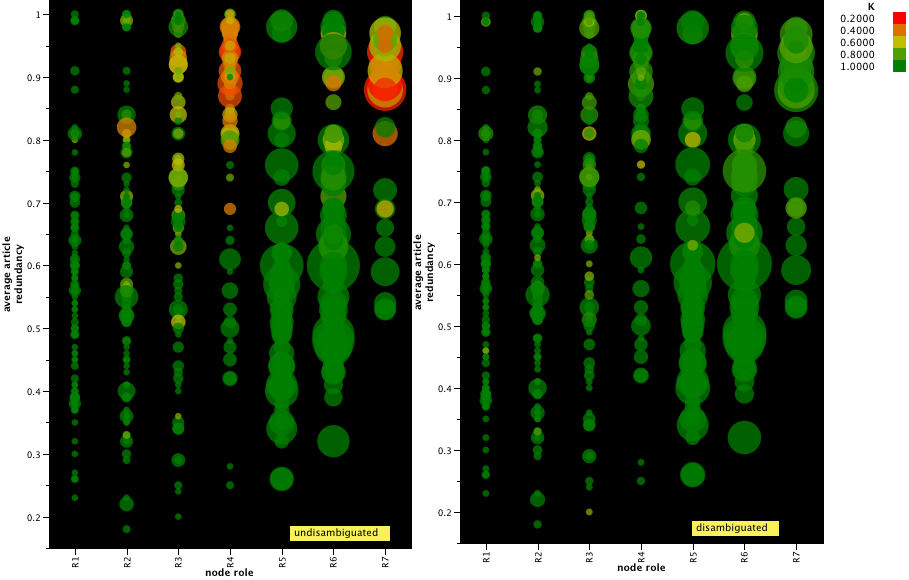, height= 10cm, width= 16cm}
\caption{Comparison of error due to homonymy based on groundtruth sample. Left: undisambiguated. Right:  disambiguated.}
\label{fig:kperformance-bubble}
\end{figure*}

Fig~\ref{fig:kperformance-bubble} visualizes for all author names in the groundtruth sample the improvement obtained by disambiguation using our algorithm. Circles represent author names, the sizes represent the number of papers coauthored by authors with that name, and the colors represent the K obtained for that author name. The figure highlights the critical gains in resolution for author names with high average article redundancies. The bubble plots underline how in particular those author names with average article redundancies between $0.85$ and $1.0$ (that correspond to the peak in fig \ref{fig:averagearticleredundancy-comparison} that we interpret as the East Asian team component in our data) suffers from lack of resolution due to name homonymy, and that the coauthor based disambiguation method studied here achieves very good improvements for that component. Only a small fraction of author names with K lower than $0.8$ remain. 

\subsection*{Resulting Changes in Network Structure}
\begin{figure}
\centering
\epsfig{file=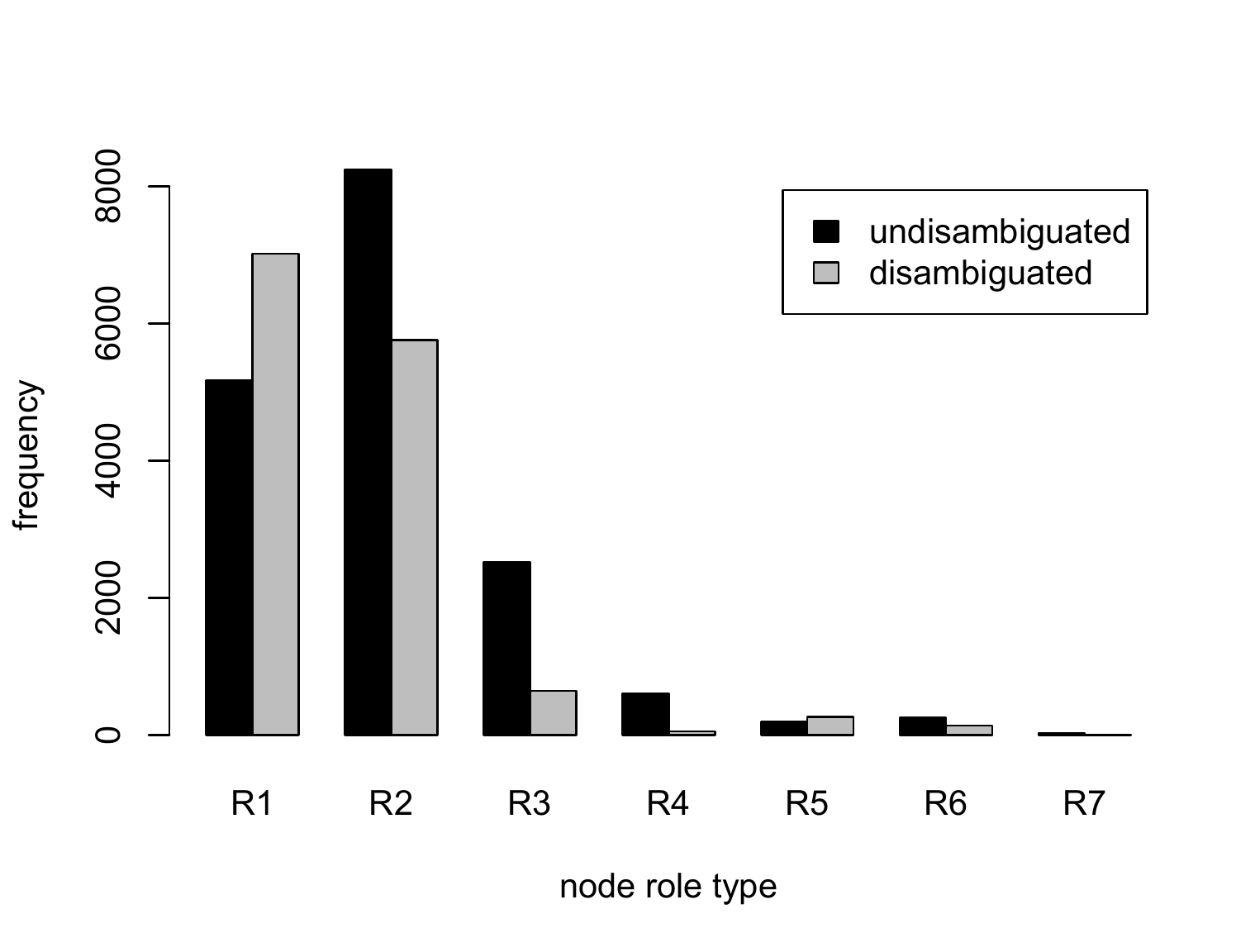, height= 6cm, width= 8cm}
\caption{Frequency distribution of node role types.}
\label{fig:noderoles}
\end{figure}

Rebuilding the co-author network from the disambiguated author data, we obtain a network almost unchanged in size, with $18,411$ nodes instead of $18,419$, but with a significantly smaller giant component of $14,057$ nodes ($76.4\%$) instead of previously $17,250$ ($93.7\%$). In fig~\ref{fig:noderoles} the node role distributions for the giant components of the undisambiguated and the disambiguated  network are compared. Disambiguation strongly affects the distribution of node role types in the entire population, reducing the proportions of the inter-cluster linking nodes R2 ($-14\%$), R3 ($-69\%$), R4 ($-89\%$), R6 ($-33\%$) and R7 ($-75\%$), and increasing the proportions of those nodes with no connections to other clusters, R1 ($+66\%$) and R5 ($+69\%$). The resulting node role distributions for hub nodes and non-hub nodes show steady declines with increasing external linking. This new result would match with the naive intuition to expect smaller and less interlinked nodes to be more numerous, since a larger node (author with more papers) and a more externally linked node (author moving between groups or participating in inter-group collaborations) requires more resources, and hence should be more rare.  Those smaller and less interlinked nodes would typically represent students publishing a couple of papers with a research group and its PI, before they move on, either to another research field or leaving science. The previous distribution, with peaks for R2 and R6 type nodes, looked counterintuitive in need for explanation for its deviation from this simple assumption.

\begin{figure*}
\centering$
\begin{array}{cc}
\epsfig{file=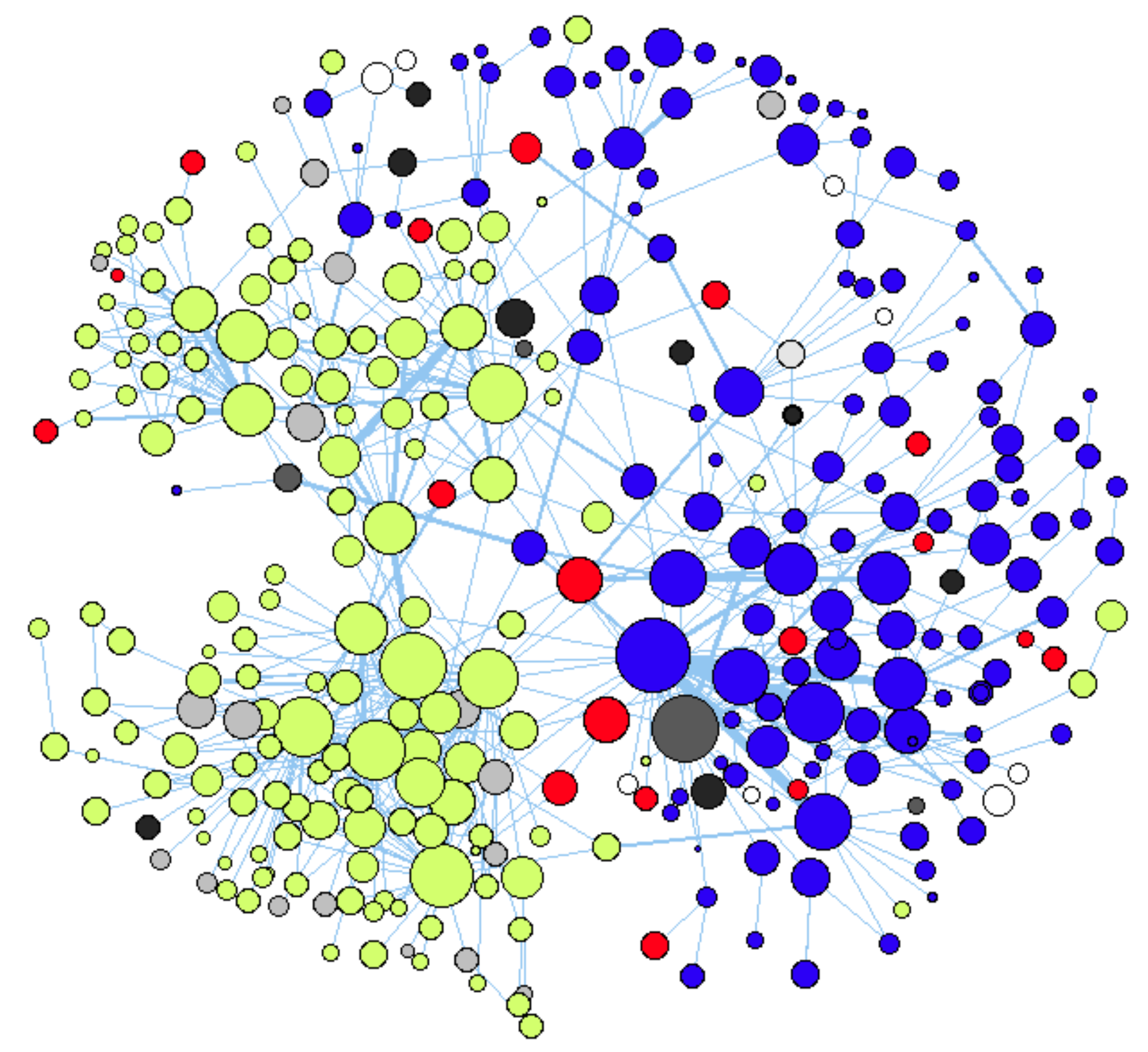, height= 7cm, width= 7cm}
\epsfig{file=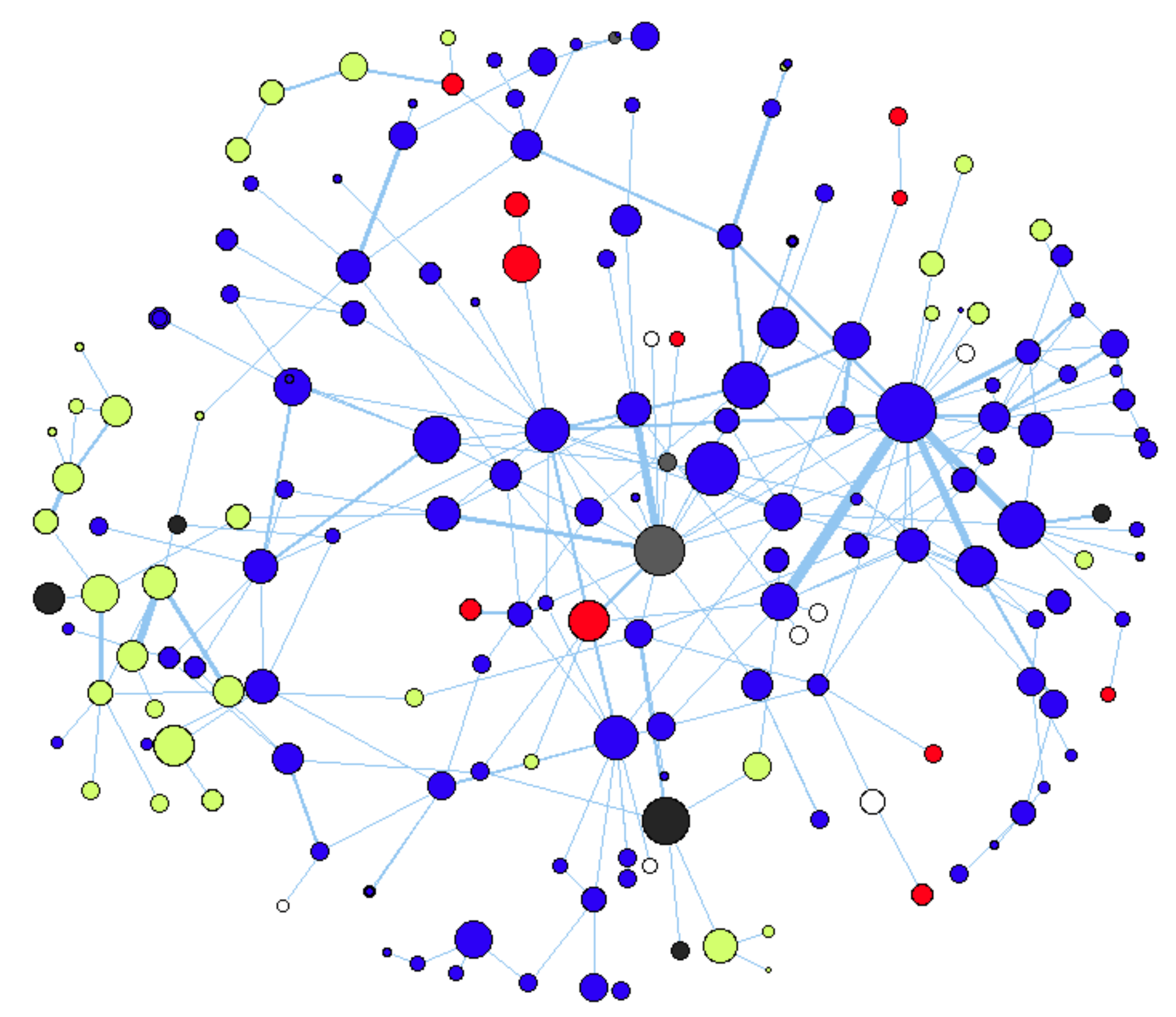, height= 7cm, width= 7cm}
\end{array}$
\caption{Collaboration Networks. Left: undisambiguated; Right: after author name disambiguation}
\label{fig:collaborationnetwork}
\end{figure*}

Fig~\ref{fig:collaborationnetwork} shows on the left hand side the global network of intensive intergroup collaboration derived in~\cite{velden2010} based on the non-disambiguated data. The right hand side in contrast depicts the network build from the disambiguated data. Note that nodes in this network represent co-author clusters, which we interpret as the footprints of collectives of closely collaborating authors. The disambiguated network is smaller in size, containing only 181 instead of 326 clusters that are less densely interlinked (node degree mean = 3.9 for the undisambiguated network vs. a node degree mean = 2.8 for the disambiguated network). We observe changes in particular in the subnetwork of Asian affiliated groups (represented by light green nodes) whose relative size shrinks from including 43\% of clusters to including only 19\% of clusters. This is presumably a consequence of the fact that, based on our groundtruth sample,  for authors with average article redundancies of $0.85$ and higher, splitting ($35\%$), weakening ($31\%$), and deletion  ($19\%$), are the dominant results of disambiguation. By contrast, disambiguation affects authors with average article redundancies below $0.85$ much less, with splitting ($9\%$), weakening ($6\%$), and deletion ($5\%$) affecting a much smaller portion of nodes in the ground truth. Further, due to the observed reduction of satellite connector nodes (R4) and global hubs (R7) in the network (nodes that, as the ground truth suggests, had particular high average article redundancies in the non-disambiguated network as shown in figure~\ref{fig:kperformance-bubble}), the interlinking of Asian affiliated groups has been drastically reduced.  

It may be worthwhile to note that for the ground truth data the major part of the remaining error in the disambiguated data can be attributed to over splitting ($15.9\%$ of names), not over merging ($2.6\%$ of names) or the combination of over merging and over splitting ($4.6\%$ of names), with similar levels of severity of error  as measured by $K$ found for over merging and over splitting. We conclude that for the algorithmic approach presented here, in spite of very good improvements, perfect resolution  remains a challenge. There is the issue of overmerging, because using co-author overlap as the central feature for disambiguation leads to the circular problem of disambiguating the co-authors first (how informative is it when on two papers the author ``LEE, H'' has a coauthor with last name ``LI'', given  that ``LI'' is an extremely common name in  our data set?). And there is the issue of over splitting from relying on continuity in co-authorship (and self-citation) alone to suggest author identity. Discontinuities in coauthor overlap arise because of e.g. team dynamics - small student dominated teams and rapid turnover of team members will imply coauthor discontinuity for the senior lead author; author mobility - career moves, joining a new group or building up new group may induce an abrupt change in coauthors; or the research focus of an author - if that focus is broad, he or she may coauthor with disjoint groups of coauthors. The influence of such factors may well differ between data sets due to their characteristics, e.g. the extent of the temporal period they cover, like in our case we analyze data covering $22$ years, allowing for numerous career movements of the authors represented.

\subsection*{Groundtruth Independent Evaluation}

\begin{figure}
\centering
\epsfig{file=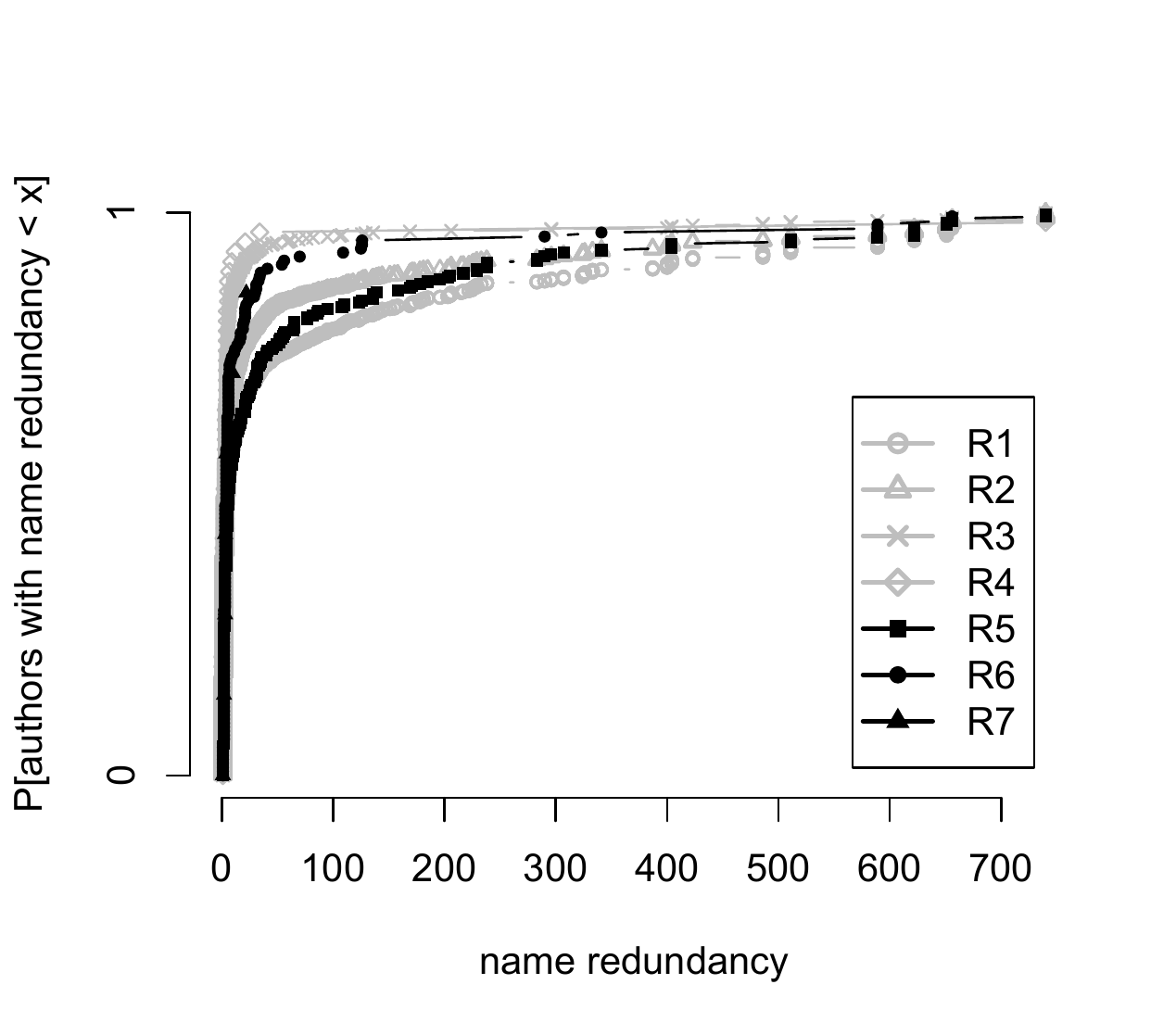, height= 10cm, width= 10cm}
\caption{Cumulative probability distributions of name redundancies
after disambiguation.}
\label{fig:nameredundancies-dis}
\end{figure}

In this section we investigate the effect of disambiguation on the role specific raw name redundancy distributions presented in fig.~\ref{fig:nameredundancies-nondis} for the undisambiguated network. If our naive expectation of a random distribution of raw name redundancies among nodes of different node role types is justified\footnote{This assumption would be violated in a situation where cultural differences in naming traditions are correlated with cultural differences in scientific collaboration, leading to different between-cluster linking patterns and consequently to differences in the distribution of name redundancies across node roles for parts of the network.}, then successful disambiguation should be reflected in close agreement between those distributions for the different role types in a network. Indeed, when we plot these distributions for the disambiguated network, fig~\ref{fig:nameredundancies-dis}, we find a dramatic change in that now all curves have become much more similar, all reflecting a long tailed distribution of raw name redundancies, but without very heavy tails. This suggests that very common names have become more equally distributed between node roles. Also in contrast to the earlier diagram, the distributions with a slightly heavier tail belong to the less outwardly linked node role types (R1, R2, and R5), indicating that some of the high redundancy names that were previously legitimately included in the strong outward linking node role types, may have suffered from over-splitting; the dominant error type remaining for our algorithm, as discussed above.
We suggest that such a node role dependent name redundancy analysis might be a very useful tool for assessing the degree of distortion in a co-author network due to name homonymy and for comparing the effect of various algorithmic disambiguation attempts.


\section{Conclusion}
We show that author name homonymy is a serious problem for the analysis of large-scale co-author networks. We derive error estimates from a ground truth sample that is statistically representative of different types of nodes in the network distinguished by their role for the connectivity of the clustered network. Those estimates confirm that a large majority of those nodes that determine the interlinking between co-author clusters in the undisambiguated network include false merges of author identities due to name homonymy. The Asian component of the global group collaboration network is most affected, due to the commonality of last names in Chinese and Korean naming traditions, the homogeneity of these traditions in a geographical area resulting in exclusively common name co-author teams, and the strong representation of Chinese and Korean groups among the Asian affiliated groups in the scientific field studied here. This explains the peculiar dense clustering observed for those groups in our previous work~\cite{velden2010}.

The disambiguation algorithm presented here deals effectively with those distortions. It rests on  a co-author overlap feature that has been found to be very effective in previous work~\cite{kang2009}. To increase performance we add self-citation as a feature, and a cut-off parameter to protect last names of low name commonality from the negative effects of disambiguation. Applying this algorithm produces significant improvements, in particular for those nodes with a critical role in inter-cluster connectivity. The great advantage of this algorithm is its scalability for large data sets and its broad applicability as it uses only a minimal set of data features (co-authors and self-citation).

We further suggest that we can gain insights on the distortion of network structure due to name homonymy without investing in the expensive creation of a ground truth sample. The distinction of classes of network nodes to reflect the mesocopic structure of a clustered network following~\cite{guimera2007} in combination with the quantification of name redundancy introduced in this work, provides a powerful lens to assess network distortion and the reduction of distortion in the network after disambiguation. The refinement and quantification of this potentially very useful analytic tool is left to future work.

\section{Acknowledgments}
This research has been made possible through financial support by the National Science Foundation through grants OCI-1025679, and  \#0404553. We are grateful to Lisa Bacis, Bryanna Gulotta, Justin Hoffman, Anna Krivolapova, David Smola,  and Adam Spar for their contribution to establishing the ground truth of the data set used in this study. 

\appendix
\section{Groundtruth Data}
We have manually established the ground truth for the author names in the node role stratified sample described in section~\ref{sec:groundtruth}. To find information on the actual identities of authors with the same combination of last name and initials, we looked up full names and institutional affiliations, if given, in the full text version of articles. We further used biographic information and affiliation information gleaned from personal homepages and institutional web pages, as well as topic information from article titles and abstracts to establish topical closeness.

Obviously, even the 'ground truth' is not necessarily the truth, because due to lack of evidence legitimate merges of identities may have been left out, and occasionally subjective judgements on topic closeness or similarity of institutional affiliation may have led to invalid merge decisions. 
\bibliographystyle{abbrv}
\bibliography{disambiguation-preprint}

\end{document}